\documentclass[10pt,letterpaper]{article}
\usepackage[top=0.85in,left=2.75in,footskip=0.75in]{geometry}
\usepackage{amsmath,amssymb}
\usepackage{changepage}
\usepackage[utf8x]{inputenc}
\usepackage[caption=false]{subfig}
\usepackage{wrapfig}
\usepackage{textcomp,marvosym}
\usepackage{cite}
\usepackage{nameref,hyperref}
\usepackage[right]{lineno}
\usepackage{microtype}
\DisableLigatures[f]{encoding = *, family = * }
\usepackage[dvipsnames]{xcolor}
\usepackage{array}
\usepackage{makecell}
\usepackage[normalem]{ulem}
\newcolumntype{+}{!{\vrule width 2pt}}
\newlength\savedwidth

\raggedright
\usepackage[aboveskip=1pt,labelfont=bf,labelsep=period,justification=raggedright,singlelinecheck=off]{caption}

\makeatletter
\renewcommand{\@biblabel}[1]{\quad#1.}
\makeatother
\usepackage{lastpage,fancyhdr,graphicx}
\usepackage{epstopdf}
\fancyheadoffset[L]{2.25in}
\fancyfootoffset[L]{2.25in}
\begin{document}
\vspace*{0.2in}

\begin{flushleft}
{\Large\bf
A Mathematical Model of Local and Global \\[0.5ex]\hspace{-0.8ex} Attention in  Natural Scene Viewing
} 
% Please use "sentence case" for title and headings (capitalize only the first word in a title (or heading), the first word in a subtitle (or subheading), and any proper nouns).

Noa Malem-Shinitski\textsuperscript{1},
Manfred Opper\textsuperscript{2},
Sebastian Reich\textsuperscript{1},
Lisa Schwetlick\textsuperscript{3},
Stefan A.~Seelig\textsuperscript{3\ddag},
Ralf Engbert\textsuperscript{3\ddag},

\bigskip
\textbf{1} Institute of Mathematics, University of Potsdam, Potsdam, Germany

\textbf{2} Department of Artificial Intelligence, Technische Universität Berlin, Berlin, Germany

\textbf{3} Department of Psychology, University of Potsdam, Potsdam, Germany

\bigskip

% Insert additional author notes using the symbols described below. Insert symbol callouts after author names as necessary.
% Remove or comment out the author notes below if they aren't used.
% Primary Equal Contribution Note
%\Yinyang These authors contributed equally to this work.
% Additional Equal Contribution Note
% Also use this double-dagger symbol for special authorship notes, such as senior authorship.
%\ddag These authors also contributed equally to this work.
% \textcurrency Current Address: Dept/Program/Center, Institution Name, City, State, Country % change symbol to "\textcurrency a" if more than one current address note
% \textcurrency b Insert second current address 
% \textcurrency c Insert third current address
% \dag Deceased
% Group/Consortium Author Note
% \textpilcrow Membership list can be found in the Acknowledgments section.
% Use the asterisk to denote corresponding authorship and provide email address in note below.
* malem@uni-potsdam.de

\end{flushleft}
\section*{Abstract}
Understanding the decision process underlying gaze control is an important question in cognitive neuroscience with applications in diverse fields ranging from psychology to computer vision. The decision for choosing an upcoming saccade target can be framed as a 
selection process between two states: Should the observer further inspect the information near the current gaze position (local attention) or continue with exploration of other patches of the given scene (global attention)? Here we propose and investigate a mathematical model motivated by switching between these two attentional states during scene viewing. The model is derived from a minimal set of assumptions that generates realistic eye movement behavior. We implemented a Bayesian approach for model parameter inference based on the model's likelihood function. In order to simplify the inference, we applied data augmentation methods that allowed the use of conjugate priors and the construction of an efficient Gibbs sampler. This approach turned out to be numerically efficient and permitted fitting interindividual differences in saccade statistics. Thus, the main contribution of our modeling approach is two--fold; first, we propose a new model for saccade generation in scene viewing. Second, we demonstrate the use of novel methods from Bayesian inference in the field of scan path modeling.

% Please keep the Author Summary between 150 and 200 words
% Use first person. PLOS ONE authors please skip this step. 
% Author Summary not valid for PLOS ONE submissions.  
\section*{Author summary}
Switching between local and global attention is a general strategy in human information processing. We investigate whether this strategy is a viable approach to model sequences of fixations generated by a human observer in a free viewing task with natural scenes. Variants of the basic model are used to predict to the experimental data based on Bayesian inference. Results indicate a high predictive power for both aggregated data and individual differences across observers. The combination of a novel model with state-of-the-art Bayesian methods lends support to our two-state model using local and global internal attention states for controlling eye movements.

% \linenumbers

% Use "Eq" instead of "Equation" for equation citations.
\section*{Introduction}
\label{sec:intro}
The human visual system acquires high-acuity information from a rather small region (the fovea) surrounding the center of gaze \cite{chalupa2004visual}. The foveal organization of the visual system has two immediate consequences. First, visual perception of natural scenes depends critically on the control of precise and fast eye movements (saccades) that move regions of interest into the fovea for high-acuity processing. During a typical visual task (e.g., scene viewing or reading), saccades occur at a rate of $3$ to $4$ per second \cite{Findlay2003}. Second, the decision process for an upcoming saccade target poses a dilemma: should the observer further exploit the information near the fovea or continue with exploration of other patches within the given scene? The latter problem is critical for scene viewing \cite{gameiro2017exploration,Ehinger2018} and relevant to the broader field of cognitive processes in knowledge acquisition \cite{Berger2014}.

Observers select saccade targets from a priority map \cite{Bisley2019} that represents objects and regions within a given scene according to their attentional weight. Over the last decades, computational modeling of visual attention for natural scenes \cite{Itti2000} resulted in a broad range of successful models \cite{Borji2012} of priority maps. These models use feature maps to combine low-level saliency and top-down control. Recently, deep neural network (DNN) models achieved state--of--the--art performances in predicting saliency maps from images \cite{Kummerer2014,Kummerer2016}. From these advances, the problem of modeling priority maps seems basically solved \cite{Einhauser2010, kummererSaliencyBenchmarkingMade2018}: for an arbitrary natural image, computational models can generate a prediction of fixation density in experiments with human observers.

\begin{figure}[h]
\begin{center}
  \includegraphics[width=0.99\textwidth]{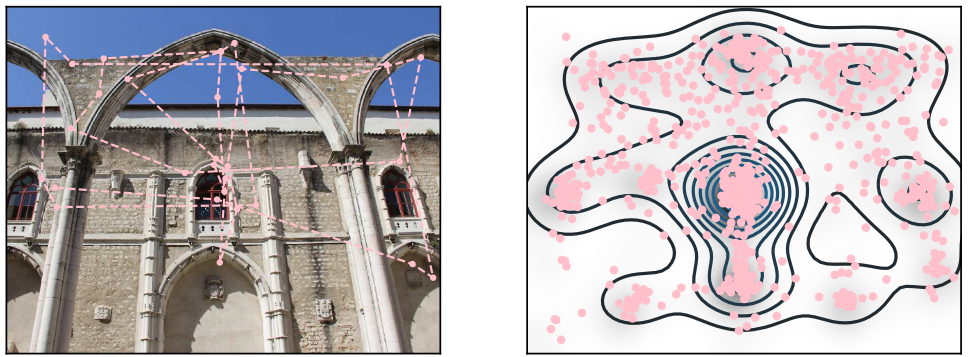}
\end{center}
\caption{{ \bf Experimental scanpath and fixations density.}
{\sl Left.} An image and a scan path. Each dot is a fixation and the dashed line illustrates the saccade. {\sl Right.} The empirical fixation density map as generated by aggregating the fixations from all subjects for a given image.}
\label{fig:data}
\end{figure}

The next step in modeling human visual behavior is fundamentally related to the fact that eye movements introduce sequential steps in information processing. Since access to visual information is effectively limited to the fovea, the full sequence of saccadic gaze shifts (scan path) needs to be modeled in order to understand the underlying principles. Understanding how human observers shift their attention while looking at an image requires quantifying the scan paths (Figure \ref{fig:data}).

So far, few models for scan path generation and prediction have been proposed. These models can be generally classified into two groups, where one group of models is hypothesis--based and the other is hypothesis--free. The second group includes models which use state of the art deep learning techniques \cite{shao2017scanpath,kuemmerer2019deepgaze3}. While
these models capture structure present in the data, they provide only very limited insights into the underlying principles of scan path generation. Another critical point for experimental research is that deep learning models require a lot of training data, which are typically unavailable for single observers. Thus, current deep learning approaches do not capture interindividual differences in statistical properties of scan paths.

Hypothesis--based models rely on cognitive and neural assumptions of human perception and oculomotor control that were derived from known biological mechanism and well-established experimental effects \cite{tatler2008systematic,Zelinsky2008,LeMeur2015,Engbert2015,Tatler2017}. Thus, the key goals of hypothesis--based models are (i) to implement these assumptions in a fully quantitative way and build a generative model, (ii) to fit the model to experimental data for hypothesis testing (statistical inference), and, finally, (iii) to provide explanations for interindividual differences in experimental data sets \cite{Schuett2017}.

In the current study, we introduce a new model which belongs to the class of hypothesis--based models. As stated above, we do not model the construction of a priority map. Rather, we address the question of how saccades are generated given a specific static priority map, and use the experimental priority map as input to our model. Our central hypothesis is that the generation of scan paths is based on switching between two internal states of local versus global attention. In this view, the generation of each saccade is a decision process, where the observer has to choose between following the local attention map, and perform a short saccade for staying in the immediate surrounding of the current fixation and the global attention map, and perform long saccade to explore a new region of the visual environment. We assume that the decision is based on the information currently available to the observer. Specifically, this assumption translates into a higher probability of following the local attention map if the ratio of priority values of the current fixated location and the previously fixated location is high. This hypothesis follows an assumption that the area next to a location with high priority also has high priority. This assumption is valid for natural images used in this work.

In implementing a model which changes between local and global attention mode, we continue the line of work that started already in $1976$ with the work of Frost and P{\"o}ppel \cite{frost1976different}. In work by Unema et al. \cite{unema2005time} and Helmert et al.  \cite{helmert2005two} it was argued that global attention mode is limited to the beginning of the scan path. Later work by Tatler et al. \cite{tatler2008systematic} showed that this is not the case. Thus our model allows the choice between a local or global attention policy throughout the entire viewing period.

Our model might also be interpreted in the context of the Exploration--Exploitation dilemma \cite{cohen2007should,berger2014exploration}. In this framework, a decision for choosing an upcoming saccade target is based on two alternatives: Should the observer further exploit the information near the current gazeposition or continue with exploration of other patches within the given scene? Hence, saccade that is generated by the local attention map corresponds to an exploitation step, and a saccade that is generated by the global attention map corresponds to an exploration step.

The idea of exploration and exploitation intentions in visual behavior was studied previously by Gameiro et al.~\cite{gameiro2017exploration}. Their work demonstrated experimentally that the tendency for exploration or exploitation, measured by saccade amplitude and fixation duration, depends on size and spatial properties of the stimulus. The characterization of the exploratory or exploitative tendencies was done using the statistics of the entire scan paths.

Different from the approach taken by Gameiro et al. \cite{gameiro2017exploration}, we analyze the individual saccades rather than entire scan paths. Our generative model tags each saccade as either a step that follows the local attention map, or a step that follows the global map.

In the current study, we model natural scene viewing which cannot be directly associated with a reward. Thus our model does not have the notion of value when choosing which policy to follow. As the terminology of Exploration and Exploitation is associated very often with the notion of the value of the decision, we avoid this terminology, and use the terminology of Local and Global Attention policies, rather than Exploration and Exploitation policies.

We aim at a minimal model to keep computations efficient and to facilitate interpretations of the model behavior, and we do not expect it to captures all the known features of human vision. A critical component of our approach is the application of Bayesian statistics to fit the model to experimental scan path data. We use the fitted model to quantify how well the model describes the experimental data. Further we test different variations of the model, which correspond to different hypotheses, to determine which hypothesis corresponds best to the experimental data.

In the next section we describe the details of our basic model and explain the computation of the likelihood function as a fundamental tool for statistical inference. We construct the model in a modular way and relate each part to one of the assumptions we would like to investigate. Next, we describe the process of fitting the model parameters to experimental data. In the Results, we compare several statistics of simulated data to the statistics of the experimental data. We also analyze different variants of the basic model and quantify how well each one of them describes the data using the model's likelihood function. We close with the Discussion of our results in the context of current problems in understanding scan path generation during scene viewing.

\section*{Materials and methods}
\label{sec:methods}
\subsection*{The Local and Global Attention for scan path generation}
\label{sec:model}
Our theoretical investigation of local and global attention in saccadic behavior is based on the implementation of a probabilistic generative model. The static viewer independent priority map for saccadic selection \cite{Bisley2019} is thought to be the combined result of early visual processing or {\sl saliency} \cite{Itti2000} and top-down cognitive control. While various models for the computation of static priority maps exist, we extend the modeling approach to the generation of scan paths for a given static saliency map. For simplicity, we use the time-averaged fixation density \cite{Engbert2015} as an approximation of the saliency of a given image.

The static saliency map is a function $s(z): \mathbb{R}^2 \mapsto \mathbb{R}^+$ with $z = (x,y)$ being a location in an image and $s(z)$ being the probability of an average viewer to fixate this location (its saliency). As mentioned above, we approximate the saliency map by the experimentally-observed fixation density and we use $s(z)$ or $s_z$ to refer to the saliency map or the fixation density of the image at location $z$.

Generally, scan paths are sequences of fixation locations and fixations duration. In this work we model only the spatial properties of gaze control. We account only for the temporal ordering of the fixations and do not model the fixation duration. In these settings, a scan path is written down as $Z = \{z_1, z_2, ..., z_t, ..., z_T\}$ with $T$ being the number of fixations in the scan path and $z_t$ being the location of the $t$th fixation.

We begin constructing our model by assuming that the saccade generation process is a second order Markov process, which means that the probability $p(z = z_t)$ of fixating on a specific location $z_t$ at time step $t$ depends only on the location of the fixation at time $t-1$ and the fixation at time $t-2$. The probability of a full scan path is written as
\begin{align}
\label{eq:scanpath}
  p(Z) = p\left(z_1\right)p\left(z_2\right) \prod_{t=3}^{t=T}p\left(z_t | z_{t-1}, z_{t-2} \right).
\end{align}

The choice of the second order Markov process reflects our hypothesis regarding the scan path generation and will become clear in the upcoming paragraphs. In principle, it is possible to construct a simpler model which corresponds to first order Markov process. This would correspond to slightly different assumptions regarding the scan path generation and we refer to such a model in the section discussing simplified models.

We describe the probability of the next fixation being $z_t$ given that the previous two fixation location were $z_{t-1}$ and $z_{t-2}$ in terms of competing local and global attention policies:

\paragraph{Local Attention} The next fixation location is chosen close to the current fixation location following a Gaussian distribution around the current fixation location with covariance $\epsilon$, normalized over the entire image. This can be written as
\begin{align}
\label{eq:exploit}
  p_{\text{local}}\left(z_t | z_{t-1} \right) = \dfrac{n\left(z_t ; z_{t-1}, \epsilon \right)}{\sum_{z'}n\left(z' ; z_{t-1}, \epsilon \right)}
\end{align}
where $n\left(z_t ; z_{t-1}, \epsilon \right)$ is a Gaussian density with mean $z_{t-1}$ and covariance $ \epsilon = \bigl( \begin{smallmatrix}\epsilon_x & 0\\ 0 & \epsilon_y\end{smallmatrix}\bigr)$.

\paragraph{Global Attention} A potential implementation is that the next fixation location is chosen randomly from the static saliency map of the image. This policy leads to very large saccade amplitudes which are known to be less probable \cite{tatler2006long}. To integrate this prior regarding the saccade amplitudes knowledge into the model -- instead of choosing the next fixation location from the the saliency map, we modulate the saliency map by a Gaussian distribution, which gives a higher weight to areas of high saliency which are closer to the current location.

This approach results in the following expression for the exploration policy
\begin{align}
\label{eq:explore}
  p\left(z_t | z_{t-1} \right) = \dfrac{s\left(z_t\right) n\left(z_t ; z_{t-1}, \xi \right)}{\sum_{z'}s\left(z'\right) n\left(z' ; z_{t-1}, \xi \right)}
\end{align}
with $\xi$ a diagonal covariance matrix similarly to $\epsilon$, $\xi_x > \epsilon_x$ and $\xi_y > \epsilon_y$.

\begin{figure}
\begin{center}
  \includegraphics[width=0.5\textwidth]{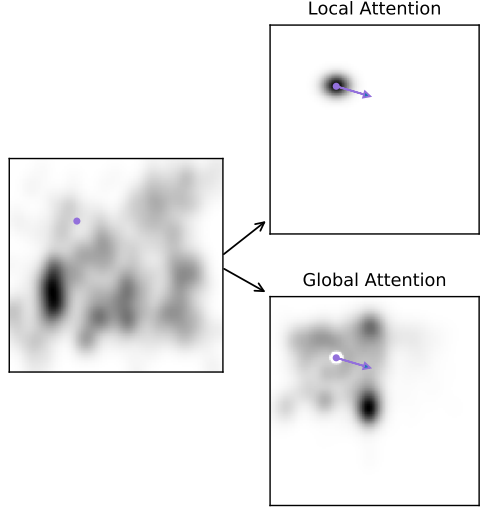}
\end{center}
\caption{{\bf Local and Global Attention maps.}
On the left is an example of an empirical saliency map, the dot indicates a fixation location. On the right are the probability maps generated by either the local attention (upper panel) or the global attention policy (lower panel). The arrow indicates a saccade.}
\label{fig:policies}
\end{figure}

Having Eq (\ref{eq:explore}) as the global attention policy may result in short saccades similar to the ones generated by the local attention policy when the current fixation is in a high priority area. A solution is to create a repulsion mechanism that forces the saccades generated by the this policy to be of at least a certain length. This is achieved by the following expression
\begin{align}
\label{eq:explore_rep}
  p_{\text{global}}\left(z_t | z_{t-1} \right) = \dfrac{\max\left(s\left(z_t\right) n\left(z_t ; z_{t-1}, \xi \right) - n\left(z_t ; z_{t-1}, \epsilon \right), 0\right)}{\sum_{z'}\max\left(s\left(z'\right) n\left(z' ; z_{t-1}, \xi \right) - n\left(z' ; z_{t-1}, \epsilon \right), 0\right)}.
\end{align}
To avoid negative values for the likelihood we take the maximum between the subtraction and 0. Figure \ref{fig:policies} visualizes the two distributions formulated in Eq (\ref{eq:exploit}) and Eq (\ref{eq:explore}).

Our assumption is that each fixation is chosen either from the local attention map described in Eq (\ref{eq:exploit}) or the global attention map described in Eq (\ref{eq:explore_rep}). This can be represented as a mixture model
\begin{align}
\label{eq:step}
  p\left(z_t | z_{t-1}, \rho \right) = \rho \: p_{\text{exploit}}\left(z_t|z_{t-1} \right) + \left(1 - \rho \right) p_{\text{explore}}\left(z_t|z_{t-1} \right).
\end{align}

The model parameter $\rho$ describes the tendency to perform a step following either the local or global attention map. It can be fixed based on prior knowledge or inferred from the experimental data. If $\rho > 0.5$ then the probability for a local step is larger than for a global step for every saccade.

Next we include in our model the assumption that $\rho$ changes depending on the fixation location. We use the notation $\rho_{t}$ to indicate that the fixation $z_t$ was generated based on $\rho_t$. Importantly, this notation does not imply that $\rho_t$ is necessarily a function of $z_t$.

We assume that the decision whether to the local or global attention maps depends on the ratio between the priority values of the current and previous fixated locations. The result is that the viewer is more likely to make a local step if the saliency value of the current fixated location is higher than the saliency value of the previous fixated location. We include this in the model with the following expression for $\rho_{t}$
\begin{align}
\label{eq:rho}
  \rho_{t} &= \sigma\left(f\left(s\right)
  \right) = \dfrac{1}{1 + \exp{\left(-f\left(s\right) \right)}}\end{align}
with
\begin{align}
  \label{eq:f}
  f\left(s\right) &= b\left(\frac{s_{t-1}}{s_{t-2}} - s^o \right)
\end{align}
with $s_{t-1} = s\left(z_{t-1} \right)$ and $b$ and $s^o$ being scalar variables.

Combining Eq (\ref{eq:scanpath}) and Eq (\ref{eq:step}), the model likelihood is written as
\begin{align}
\label{eq:like}
  p\left(Z| \Theta \right) = p\left(z_1\right)p\left(z_2\right) \prod_{t=3}^{t=T} \left( \rho_{t} \, p_{\text{local}}\left(z_t|z_{t-1} \right) + \left(1 - \rho_{t} \right) p_{\text{global}}\left(z_t|z_{t-1} \right)\right)
\end{align}
with model variables $\Theta = \{\epsilon, \xi, b, s^o\}$. Here, we chose to sample the first and second fixation from the empirical static saliency map such that $p\left(z\right) = s\left(z\right)$.

\begin{figure}[h]
\begin{center}
  \includegraphics[width=0.99\textwidth]{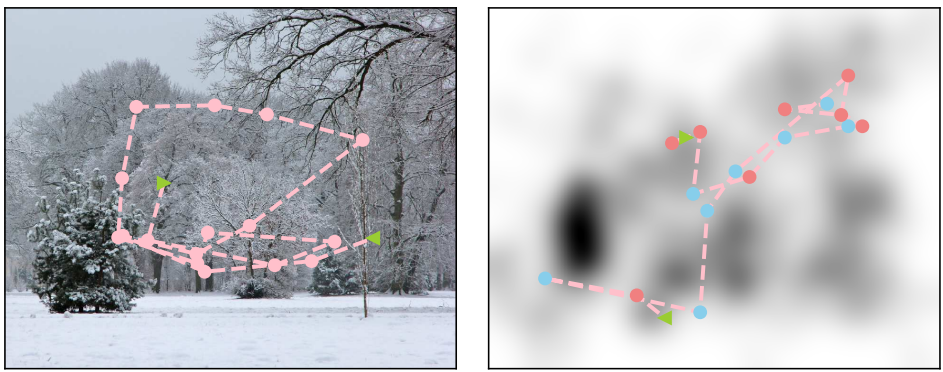}
\end{center}
\caption{{\bf Experimental and simulated scan paths.}
{\sl Left.} An image and a scan path recorded from a human observer. {\sl Right.} The experimental static saliency map and a scan path generated by the Local and Global Attention Model. The green arrow pointing right represents the second randomly selected fixation location $z_2$. The green arrow pointing left represents the last fixation in the scan path. The blue dots are fixations that were generated from an exploration step and the pink dots are fixations that were generated from an exploitation step. The experimental data shows clearly the phenomenon of saccadic momentum which is not captured by the model. This is further discussed in the Results and Discussion.}
\label{fig:scanpaths}
\end{figure}

Figure \ref{fig:scanpaths} presents a scan path generated by our model given a particular saliency map, along side a scan path recorded experimentally from a viewer viewing the image corresponding to the saliency map.

\subsection*{Simplified models}
\label{subsec:simp}
To test the different assumptions behind our full model described above, we construct three simpler models and compare their performances to the performance of the full model in the Results. To construct the models we remove one by one the assumptions on which the model is based. This results in the following competing models:

\paragraph{Local Choice Model:} Eq (\ref{eq:rho}) describes the assumption that the decision between two attentions maps depends on the ratio between the priority value of the current fixation location and the priority value of the previous fixation location. A competing assumption would be that the decision depends only on the priority value of the current fixation location. In this case we keep the model the same and only change $f\left(s\right)$
\begin{align}
  f\left(s\right) = b\left(s_{t-1} - s^o \right).
\end{align}

\paragraph{Fixed Choice Model:} We test the assumption that the decision between the modes does not depend on the saliency value of previous fixation. In this simplification of the model, rather than having $\rho_t = f\left(z_{t-1}, z_{t-2} \right)$ we have a fixed probability to chose each policy with $\rho_t = \rho$.

\paragraph{Local Saliency Model:} Last, we challenge the approach of two competing modes. In this variation of the model, each fixation is generated from a modulation of the empirical saliency map with a Gaussian around the current fixation location. This corresponds to the following fixation location likelihood
\begin{align}
\label{eq:loacl_sal}
  p\left(z_t|z_{t-1} \right) = \dfrac{s\left(z_t\right) n\left(z_t|z_{t-1}, \xi \right)}{\sum s\left(z'\right) n\left(z'|z_{t-1}, \xi \right)}
\end{align}

In the next section we describe the inference process of the full Model. As the three models described above are simplified versions of the full model we do not describe their corresponding inference processes as they can be easily derived from the inference of the full model.

\subsection*{The inference process}
\label{sec:inf}
Our approach is based on experimental results and we derive the model parameters from observed data in a Bayesian framework. This approach allows us to include prior knowledge regarding the different model parameters based on known spatial features of scan paths. It also allows us to obtain distributions over the model parameters, rather than point estimates, and to compare different variations of the model via the respective test--data likelihoods.

In the previous section we defined the likelihood of the data. Next, we describe the data augmentation methods which allow us to identify conjugate priors and construct an efficient Gibbs sampler \cite{geman1987stochastic} using the full conditional distributions over the model parameters. 

The idea behind data augmentation \cite{liu2008monte} is adding latent variables to the model, which can be considered as unobserved data, in a way that simplifies the inference of the parameters of interest. We use the standard approach and augment the Local and Global likelihood by
\begin{align}
\label{eq:qug_gamma}
  p(Z, \Gamma| \Theta) &= p(z_1)p(z_2) \prod_{t=3}^{T} p_{\text{local}}\left(z_t|z_{t-1} \right)^{\gamma_{t}} \: p_{\text{global}}\left(z_t|z_{t-1} \right)^ {1 - \gamma_{t}}
 \end{align}
with
 \begin{align}
  \label{eq:ber}
  \gamma_{t} &\sim \text{Bern}\left(\rho_{t} \right) = \text{Bern}\left(\sigma \left( f\left(s\right)\right) \right)
\end{align}
and marginalizing over $\Gamma$ results in Eq (\ref{eq:like}). 

The augmentation defines a modified generative process for the model. At each time step a variable $\gamma_t$ is drawn from a Bernoulli distribution with bias $\rho_{t}$. If the result is $1$ then the next saccade is generated following the local attention mode. If the result is $0$, the saccade is generated from the global attention mode. This construction reflects our assumption regarding the cognitive process underlying scan path generation, where each saccade follows either local or global attention mode.

 For a simple two--component mixture model with normal distribution, the augmentation described above would have been sufficient for the derivation of a Gibbs sampler \cite{gelman2013bayesian}. As the model we constructed is more complex, we need to handle the sigmoid link--function in Eq (\ref{eq:rho}) and the non-trivial form of the Global Attention distribution in Eq (\ref{eq:explore}).

With the Sigmoid function in Eq (\ref{eq:rho}) there is no straightforward way to define conjugate priors for the parameters $b$ and $s^o$ which are needed for a Gibbs sampler. To achieve conditional probabilities which are easy to sample from, we augment the model with another set of latent variables $w_t$, which follow a P\'{o}lya-Gamma distribution
\begin{align}
  \label{eq:aug_polly}
  w_t \sim \text{P}G\left(1, -f\left(s\right) \right).
\end{align}

As described in \cite{polson2013bayesian} for the case of logistic regression, the usage of this augmentation scheme results in conditional distributions for $b$ and $s^o$ which are Gaussian and can be sampled from easily. The full derivation of the discussed conditional distributions can be found in the supplementary material.

After adding the two sets of latent variable we can define conjugate priors for the parameters:
\begin{align}
  \epsilon_{x/y} &\sim \text{IG}\left(\epsilon_{x/y} ; \alpha_{\epsilon_{x/y}}, \beta_{\epsilon_{x/y}} \right)\\
  \xi_{x/y} &\sim \text{IG}\left(\xi_{x/y}; \alpha_{\xi_{x/y}}, \beta_{\xi_{x/y}} \right) \\
  b &\sim \mathcal{N}\left(b; \mu_b, \sigma_b \right) \\
  s^o &\sim \mathcal{N}\left(s^o ; \mu_{s^o}, \sigma_{s^o} \right)
\end{align}
where $\text{IG}$ is the Inverse Gamma distribution, and $\mathcal{N}$ is the Gaussian distribution.

The prior distributions described above include hyperparameters. These parameters were chosen and not inferred from the data. The hyperparameters related to the prior distributions over $\epsilon_{x/y}$ and $\xi_{x/y}$ were chosen based on known characteristics of human saccades, such as that typical saccade amplitudes range from $0.5$ and up to $40$ visual degrees \cite{wong2014eye}. The hyperparameters related to $b$ and $s^o$ were chosen to be on the same scale of the average $\frac{s_{t-1}}{s_{t-2}}$ from the data. Further, all of the hyperparameters were chosen to induce wide prior distributions. 

Combining the likelihood in Eq (\ref{eq:like}) with the priors defined above, the posterior distribution over the model parameters and the latent parameters is given by
\begin{align}
\label{eq:posterior}
  p\left(\Theta, \Gamma, W | Z \right) &\propto p\left(Z|\Theta, \Gamma , W\right) p\left(\Gamma | \Theta \right) p \left(W|\Theta \right)p\left(\Theta \right)
\end{align}
with
\begin{align*}
  p\left(\Theta \right) &= p\left(\epsilon \right) p\left(\xi \right) p\left(b \right) p\left(s^o \right)
\end{align*}.

We can sample easily from the conditional distributions of $b$ and $s^o$. This is not the case for $\epsilon$ and $\xi$ because of the form of $p_{\text{explore}}$.

Due to the complex form of the global attention expression in Eq (\ref{eq:explore_rep}), which includes both $\xi$ and $\epsilon$, there is no closed form for the conditional distribution of these parameters. Thus, we resort to a technique known as MCMC within Gibbs \cite{gilks1992adaptive,martino2015fast} and in each iteration of the Gibbs sampler we evaluate the conditional distributions of $\xi$ and $\epsilon$ using an Hybrid Monte Carlo step \cite{duane1987hybrid}, also known as Hamiltonian Monte Carlo.

For further technical details regarding the augmentation and the HMC sampler please see the supplementary material.

The implementation of the models and inference can be found under https://github.com/noashin/local\_global\_attention\_model .

% Results and Discussion can be combined.
\section*{Results}
\label{sec:results}

In this work we propose a Global and Local Attention Model for scan path generation. In the previous section we derived the model equations from the basic two modes approach. We described the inference process of our model when applied to experimental data. In this Section, we present the results of the inference process. First, we analyzed the reliability of our procedures by fitting the model to artificial data generated from the model with known parameter values. Next, we fit the model to the experimental data and test the statistics of the data generated from the model against the experimental data. Finally, we quantitatively compare different versions of the model.

\subsection*{Model parameters estimation}
As presented in the Methods Section, the inference process includes using an MCMC approach to evaluate the posterior function over the model parameters. This approach is exact in the limit of an infinite number of samples but as we can only use a finite number of samples the result is an approximation of the actual posterior. The distribution of the inferred parameters should concentrate around their real values.

When fitting the model to experimental data it is impossible to know the real values of the model parameters as they do not relate directly to any measurable features of the data. Thus, in order to assess the performance of the inference we use data simulated by the model, in which case we know the exact values used to generate the data. If the inference process is correct we expect the resulting posterior distribution to be concentrated around the ground truth values.

We generated data from our model with the parameter values that were inferred from the experimental data. In order to see whether the inference process will have reasonable results when fitting the experimental data, the size of the generated data set is comparable to the size of the experimental data for one subject.

\begin{figure}[!h]
\begin{center}
  \includegraphics[width=0.99\textwidth]{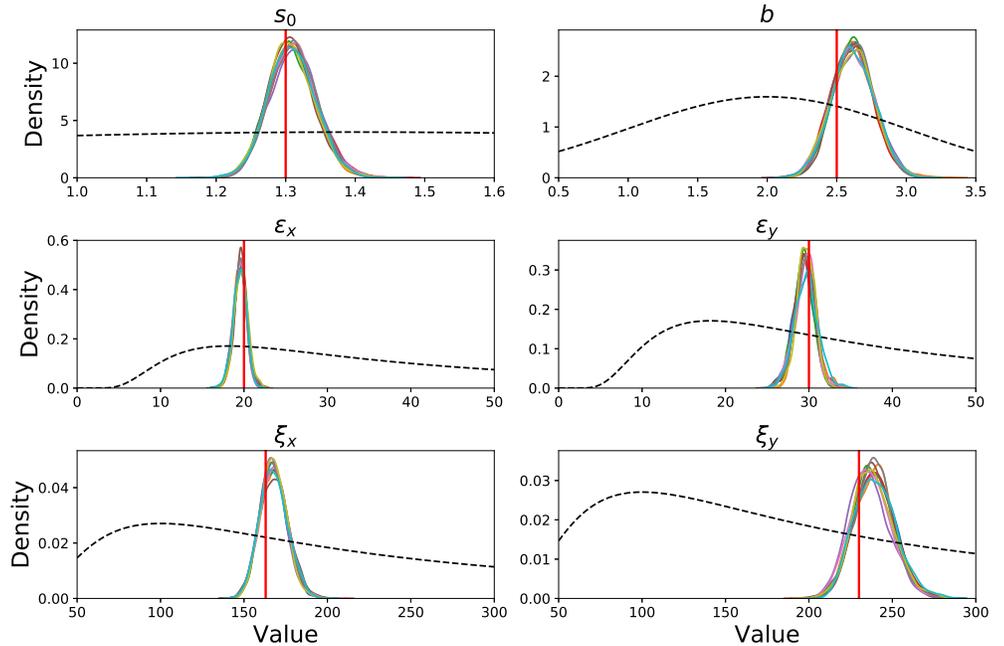}
\end{center}
\caption{{\bf Inference results on simulated data.}
Model parameter recovery. To test the inference algorithm we fit the model to simulated data with known parameters values. Each panel includes the inferred posterior distribution of each parameter after the inference process. The ten curves present 10 different inference processes starting from different values. The vertical lines are the values with which the data was generated. The black dashed curve is the prior distribution. The plotted densities are not normalized.}
\label{fig:dists}
\end{figure}

Figure \ref{fig:dists} presents the distribution over model parameters as results from the inference process with data generated by the model. Each of the ten colored curves represents a different inference process started at a different point. As expected all the curves from different runs are similar in shape. The black dashed curves present the prior distribution over the parameters. To test the model we chose the prior distribution so their modes do not overlap with the values used in the data generation. As expected the mode of the inferred parameter distribution is close to the real values used in the data generation which are noted by the vertical solid line.

We tested the model on generated data that have similar properties to the experimental data. Generated scan paths had lengths similar to the lengths of scan paths recorded experimentally. This could have the result that the generated data does not have sufficient information regarding the underlying model parameters and it explains the deviation of the distribution mode from the true parameter values.

\subsection*{Model performance on experimental data}
Our model was derived from a set of hypotheses regarding the cognitive process of saccade generation. In order to test the validity of the model, and of the corresponding hypotheses, we fit the model to the data, simulate new data using the model and check whether the features of the simulated data correspond to the features of the experimental data.

The data set used here includes the scan paths of thirty five human observers performing a memorization task over thirty natural images. The same data set was used before to evaluate other scan path models \cite{Engbert2015,Schuett2017}. The participants were presented with an image for 10 seconds and were instructed to explore the scene for a later memory test. The acquisition of the data was carried out in accordance with the Declaration of Helsinki, and informed consent was obtained for experimentation by all participants. Data from three subjects were excluded as the inference process did not converge. The data can be found under https://osf.io/me2sh/.

We fit a separate model for each subject, while using the same prior hyperparameters for all fitted models. We want to test whether the model captures subjects' tendencies that generalize over images. We use the k-fold cross-validation method with $k = 5$. All the reported quantitative results in this section are obtained from the test data averaged over the different folds.

\subsubsection*{Saliency map recovery}
\begin{figure}[!h]
\begin{center}
  \includegraphics[width=0.99\textwidth]{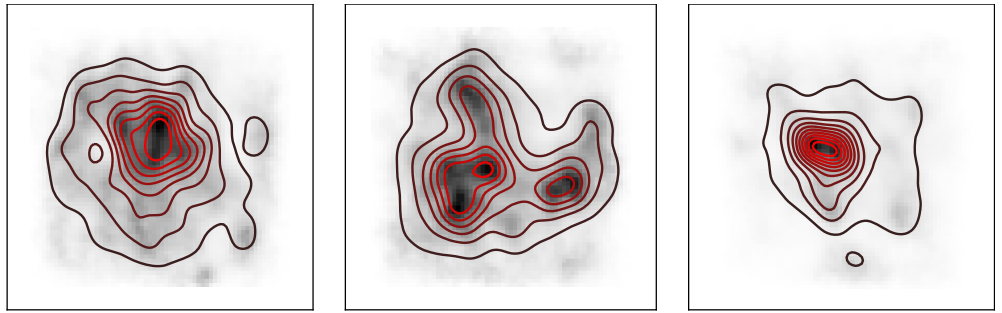}
\end{center}
\caption{{\bf Comparison between the empirical saliency map and the fixation density of data generated by the model, for three different images from the test-set.}
The empirical saliency is represented by the shading, and the contour lines represent the density of the data generated by the model. The generated data recovers the original empirical saliency map.}
\label{fig:sal_rec}
\end{figure}

Since the goal of our model is to produce a scan path for a given saliency map, the model needs to recover the empirical saliency map from experimental data. Figure \ref{fig:sal_rec} presents the comparison between empirical saliency maps and the fixation locations density of data generated by the model. We used three different empirical saliency maps from the test-set for simulation of the full Local and Global Attention model and generated data from all the models fitted to the different subjects. The contour plot includes the density of the aggregated data, and the density is the empirical saliency map. Qualitatively, as expected, the fixation density of the data generated by the model, matches the empirical saliency maps.

\subsubsection*{Saccade amplitude}
The Local and Global Attention Model was designed to capture the different saccade amplitudes generated by subjects while observing an image in a free viewing task. To estimate the model's performance we compare the amplitudes of the empirical saccades with the amplitudes of the saccades generated by the model. The comparison is done both at a population level and for each subject separately.

\begin{figure}[!h]
\begin{center}
  \includegraphics[width=0.99\textwidth]{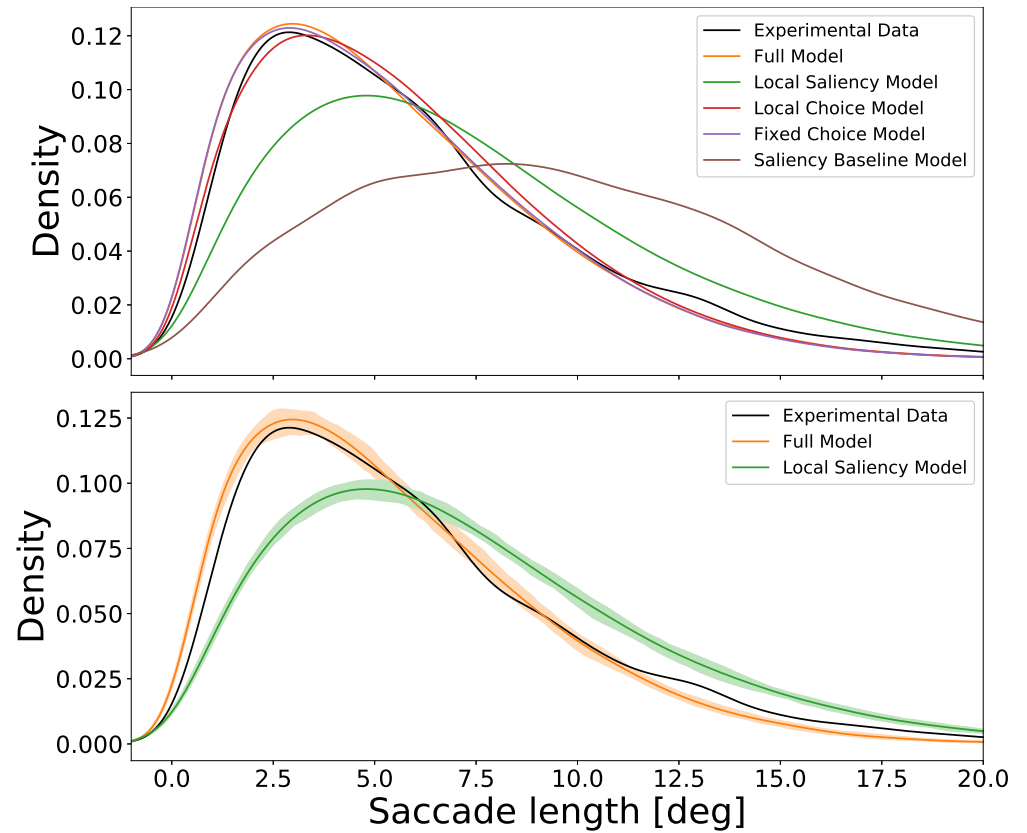}
\end{center}
\caption{{\bf Saccade amplitude density - experimental and simulated.}
Saccade amplitude density, aggregated over the data from all participants, of the experimental data and data generated by the full model and the simplified competitor models. {\sl Top.} Comparison of all models. {\sl Bottom.} Comparison between the full model and the Local Saliency Model. The shading corresponds to confidence bounds regarding the estimate of the model parameters. The full model captures the different kinds of saccade lengths, whereas the simpler models fail to do so.}
\label{fig:sacc_len}
\end{figure}

Figure \ref{fig:sacc_len} compares the empirical saccade amplitude density with the saccade amplitude density of the scan paths that were generated by the full Local and Global Attention Model and the simplified versions presented previously. The density presented is over the entire population of subjects. The black curve presents the empirical data. The orange curve corresponds to data generated by the full Local and Global Attention Model, and the other curves correspond to the different simplified models. 

As a baseline we include the Saliency Baseline Model, where the scanpath is sampled from the saliency map. As this model does not have any constraints on the saccade amplitude, other than the distance between high saliency areas in the image, the generated saccades have a much higher amplitude than the experimental data. Limiting the saccade amplitude as in the Local Saliency model by assuming a local attentional focus, results in saccades with much more realistic amplitudes.
Figure \ref{fig:sacc_len} shows that the full model performs better than the simplified models. The three simplified models tend to capture the mean saccade amplitude rather than the full variety of saccade amplitudes displayed in scan paths. This behavior is expected from the Local Saliency Model, which includes only one type of characteristic saccade amplitude whereas the full Local and Global Attention Model has two characteristic saccade amplitudes that correspond either to the local or global attention mode. 

The Bayesian inference process presented in Methods Section results in a distribution over the possible values of the model parameters. This corresponds to uncertainty regarding the values of the model parameters. The shading around the generated data curves in Figure \ref{fig:sacc_len} corresponds to this uncertainty. We sampled 50 different values from the posterior distribution of each one of the model parameters and used this configuration to generate one data set. We splitted the experimental data into training and test sets three times and repeated the fitting of the models on each training set separately, resulting in a 5-fold cross-validation. This process applies to all the results presented, unless stated otherwise. Figure \ref{fig:sacc_len} presents the result of one such training and test split. The shading represents the $95\%$ intervals around the mean density over the different data sets.

The confidence bounds are rather narrow and the density distributions of the two models are highly separable. This is a good indication that the Bayesian parameter inference is reasonable -- the saccade amplitude density does not change dramatically with the parameter configurations sampled from the posterior distributions. The confidence bounds for the Local and Fixed Choice Models behave in a similar way and are not included in the figure for clarity purposes.

\begin{figure}[!h]
\begin{center}
  \includegraphics[width=0.8\textwidth]{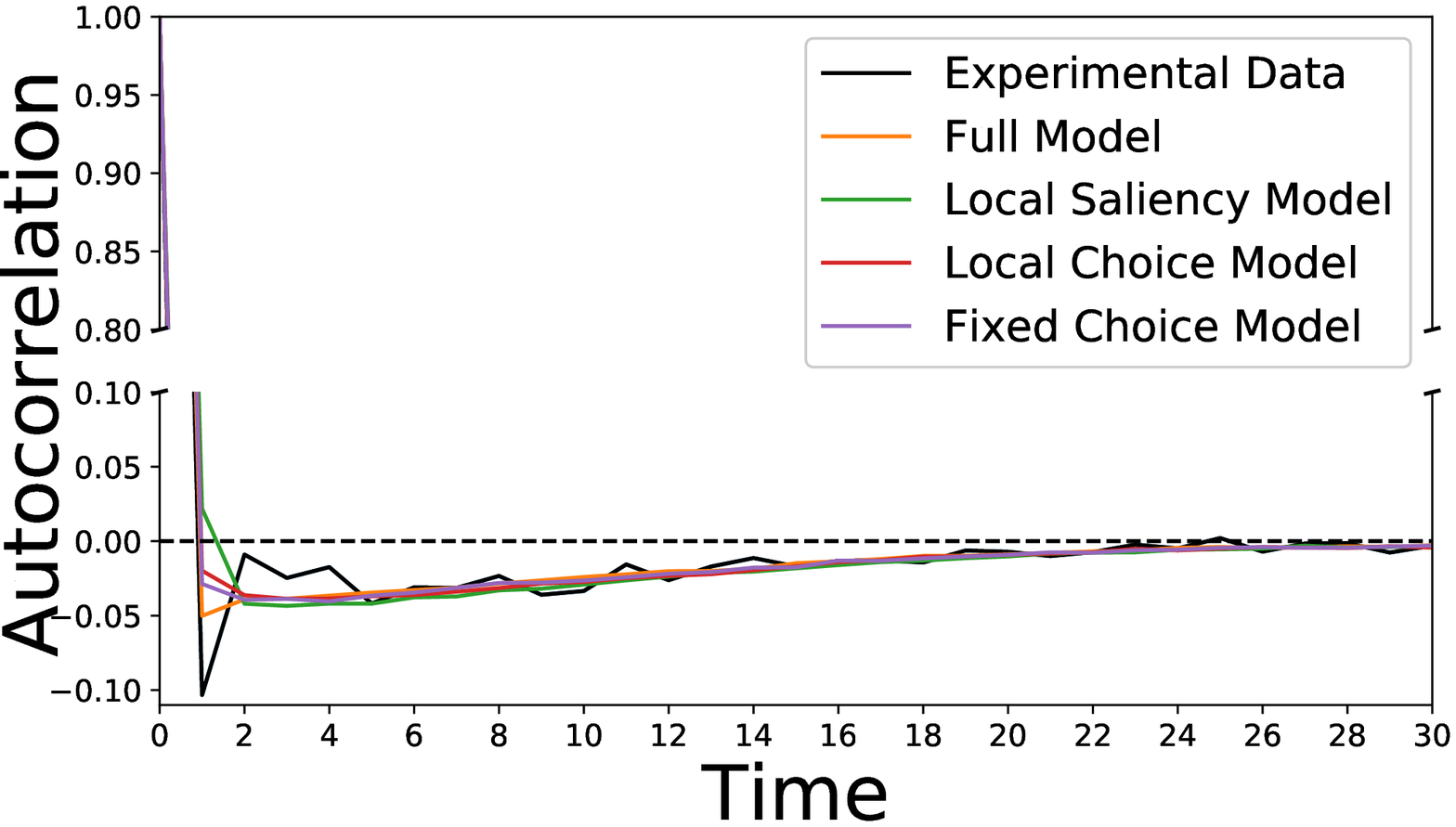}
\end{center}
\caption{{\bf Saccade amplitude autocorrelation - experimental and simulated.}}
Saccade amplitude autocorrelation, averaged over experimental data from all participants and over all simulations generated by the full model (and the various competitor models). The full Local and Global Attention Model approximates the autocorrelation in amplitude of successive saccades, whereas the simpler models fail to reproduce the lag-1 anti-correlation.
\label{fig:sacc_len_corr}
\end{figure}

The model presented in this work generates a sequence of saccades, rather than independent saccades. Thus, we would expect to see some correlation between the generated saccades. Figure \ref{fig:sacc_len_corr} presents the mean aurocorrelation of the saccade amplitude along a scan path. The experimental data shows a clear anti-correlation between the amplitude of subsequent saccades at lag 1. Thus, a short saccade is likely to be followed by a long saccade and vice-versa. Although not as strong as in the experimental data, this effect is captured by the Local and Global Attention Model. This result is expected from our modeling assumptions, since when generating fixation $z_t$ the full model has information regarding the saliency of fixation $z_{t-2}$, whereas the competing models do not have access to this information. In addition ot this lag-1 effect, it is important to note that the our model also approximates the autocorrelation function for lags up to 20.

\begin{figure}[!h]
  \includegraphics[width=0.99\textwidth]{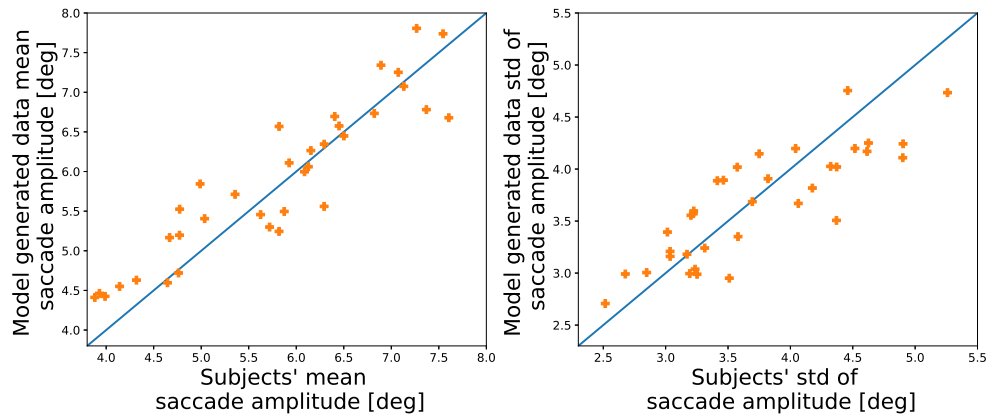}
\caption{{{\textbf{Comparison between experimental and simulated subjects' mean  and standard deviation of saccade amplitude.}}}
{\sl Left.} Participants' mean saccade amplitudes compared with the mean saccade lengths of data generated by the Local and Global Attention model. {\sl Right.} The standard deviation of the subjects' saccade amplitude compared to the standard deviation of the data generated by the Local and Global Attention model. Overall the model captures the both the mean and the standard deviation of the saccade amplitude of the different subjects.}
\label{fig:means_stds}
\end{figure}

As described above, we fit a model for each subject individually. Thus, we can investigate how well the Local and Global Attention Model captures the difference between the subjects. In the left panel in Figure \ref{fig:means_stds} we compare the mean saccade length of the empirical data and data generated from the full model for each subject. Each data point is one subject and the diagonal curve is the identity line. The presented data is from one fold of the k-fold cross validation.

Overall, the model captures the different mean saccade length of the different subjects. Not only does the model capture the different mean saccade amplitudes of the subjects, it also captures the difference in the variability of saccade amplitudes of the subjects (see the right panel of Fig.~\ref{fig:means_stds}, where the standard deviations of the saccade amplitudes are plotted per participant.)

In Table \ref{tab:r2s} we report the coefficient of determination between the mean and standard deviation of the subjects' data and of the data generated by the full Local and Global Attention Model and the competing simplified model. The coefficient of determination was averaged across the different train-test splits in the cross validation. Other than the Local Saliency model, all model perform similarly well and capture the mean and standard deviation of the saccade amplitudes of the different subject. This result indicates that the assumption of two length scales generated by local and global attention states represents a major improvement in the model fit.

\begin{table}[!h]
\centering
\begin{tabular}{ccccc}
\noalign{\smallskip} \hline \hline \noalign{\smallskip}
& \thead{Local and Global\\Attention} & \thead{Local\\Choice} & \thead{Fixed\\Choice} & \thead{Local\\Saliency} \\
\hline
$R^2$ saccade amplitude mean
&$0.93$ 
& $0.93$ 
& $0.93$
& $0.47$ \\
\hline
$R^2$ saccade amplitude std
&$0.85$ 
& $0.847$ 
& $0.85$
& $0.3$ \\
\noalign{\smallskip} \hline \noalign{\smallskip}
\end{tabular}
\caption{Comparison of the coefficient of determination, between the mean (or std) of the subjects' saccade amplitudes and the saccade amplitudes of the data generated by the different models. Other than the local saliency model, all models capture both the mean ant the standard deviation of the saccade amplitudes of the different subjects.}
\label{tab:r2s}
\end{table}

\subsubsection*{Saccade direction}

Generally, saccades can be seen as vectors characterized by amplitude and direction. After analyzing the model performance with regard to the saccade amplitude, we turn to analyze the model performance with respect to the saccade direction. 

\begin{figure}[!h]
\begin{center}
  \includegraphics[width=0.99\textwidth]{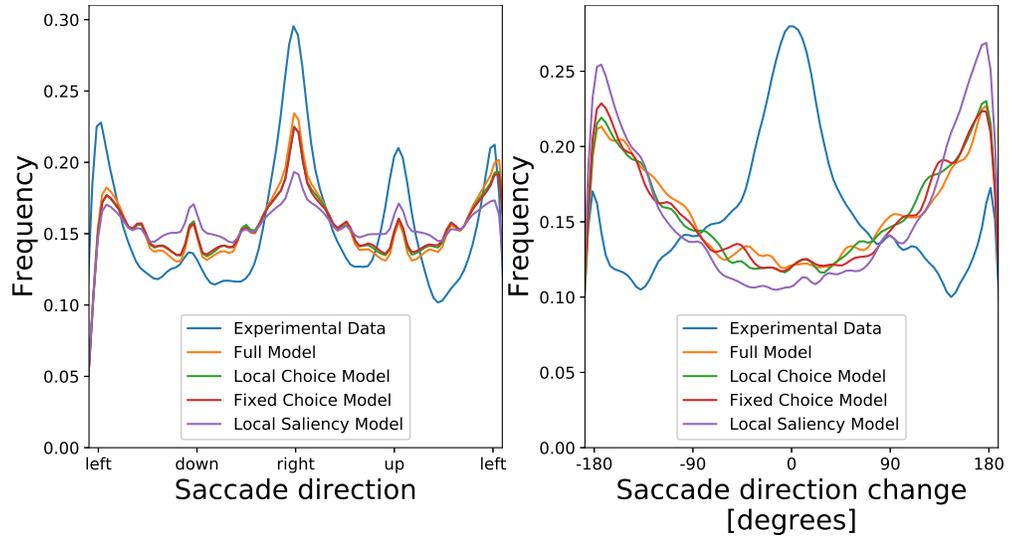}
  \end{center}
\caption{{\text{\bf Experimental and simulated saccade direction and saccade direction change frequency.}}
{\sl Left.} Absolute saccade direction. The empirical data demonstrate a strong tendency to saccades towards the left and right directions, and a weaker tendency to perform saccades directed upwards. The generated data captures the tendency to perform horizontal saccades, but not vertical saccades. {\sl Right.} Change in saccade direction. The generated data demonstrates the tendency to persist in the same direction. The models fail to capture this persistence.}
\label{fig:dirs}
% \vspace{-20pt}
\end{figure}

There are two important aspects regarding saccade direction, i.e., absolute saccade direction and the direction relative to the previous saccade. In the left panel in Figure \ref{fig:dirs} we compare the saccade direction density, over the entire population of subjects, of the empirical data and of data generated by the fitted full model and its variations. The empirical data demonstrate clear preference for horizontal saccades and a weaker tendency towards vertical upward saccades. The different variations of the model generate similar distributions of saccade directions. The data generated by the models correspond to a tendency to perform horizontal saccades, but this tendency is not as strong as in the empirical data. The empirical tendency towards vertical saccades is not captured at all by the models.

The fact that the different models capture only one preferred horizontal direction is not surprising. It is common across all the variations of the model that at each step the next fixation is generated from an ellipsoidal distribution, which has only one preferred direction. In the Discussion, we suggest variations of the model which could capture more than one dominant saccade direction.

The right panel in Figure \ref{fig:dirs} presents the frequency of the in saccade direction change values. The experimental data is characterized by a large peak around $0$ which is an indication of persistence of the current saccade direction  \cite{ritter1976evidence,breitmeyer1982existence,wilming2013saccadic}, also known as saccadic momentum. Additionally, there is a weaker peak around $180$ and $-180$ degrees which indicates a tendency to return to the previous fixated location. All models discussed here fail to reproduce this effect. The peak in the saccade direction change is only around $180$ and $-180$ degrees which is due to the hard constraints given by the image boundaries.

\subsubsection*{Model comparison}
\label{sec:model_comp}
Last, we would like to compare the performance of the full Local and Global Attention Model to the simplified variants of the model presented in the Methods Section. As measurements of the model performance we use the Receiver Operating Characteristic (ROC) and the respective Area Under the Curve (AUC) and the Normalized Scan path Saliency (NSS). These methods are widely used in the field of attention modeling and have been successfully adapted to scan path modeling \cite{peters2005components,wang2011simulating,borji2012state,riche2013saliency,kummererSaliencyBenchmarkingMade2018,kuemmerer2019deepgaze3}.

\begin{figure}
  \begin{center}
    \includegraphics[width=0.48\textwidth]{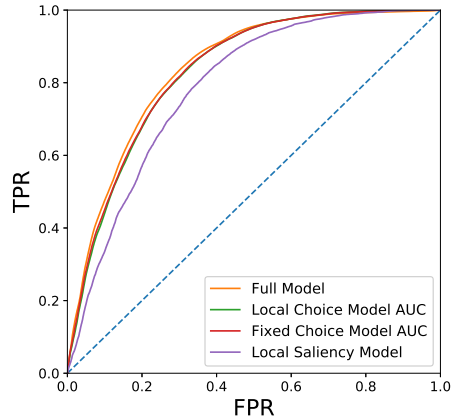}
  \end{center}
  \caption{{\bf ROC curves of the different model variants.}
The full model performs slightly better than the Local and Fixed Choice models. The Local Saliency Model performs significantly worse than the other model variants.}
\label{fig:roc}
\end{figure}

The AUC measure is a very common tool to analyze the performance of a probabilistic classifier by looking at the trade-off between True Positive and False Positive Rates (TPR and FPR), when using different thresholds for classification. In the context of attention modeling, the fixation locations are used as samples with positive labels, and as samples with negative label we used image coordinates that were sampled uniformly in the image space. This method is denoted as AUC-Borji \cite{bylinskii2018different}. In our analysis, the likelihood of the model at each time step is used as the probabilistic classifier. Figure \ref{fig:roc} presents the ROC curves for the different models. We will discuss these results shortly after presenting the NSS measure.

The NSS is the average normalized saliency (with mean zero and standard deviation of one over the image) along the scan path of fixated locations. In this work we used the likelihood value of each fixation in the scan path, and normalized the likelihood over the entire image, as one would do with a saliency map. Table \ref{tab:roc_nss} presents the AUC and NSS scores for the different model variants.

As done in the previous analysis, scores are reported on the test data-set, which was not used in fitting data, and averaged over the different samples form the posterior distributions and the folds of the cross validation process. Due to the split of the data into training and test sets, the discussed measures are sensitive to the model complexity. If a model is too complex (usually manifested by having a lot of parameters) the model will achieve high AUC and NSS over the training set but may suffer from over fitting and perform poorly on the test set.

From the results in Table \ref{tab:roc_nss} we see that the full model performs better than the other variants, and achieves higher AUC and NSS scores. Although the scores of the full model are the highest, they are only slightly better than the ones of the Local and Fixed Choice models. It is clear that the Local Saliency model performs poorly, which emphasizes the importance of the two characteristic length scales, which are present in all of the model variants other than the Local Saliency Model.

\begin{table}[!h]
\centering
\begin{tabular}{ccccc}
\noalign{\smallskip} \hline \hline \noalign{\smallskip}
& \thead{Local and Global\\Attention} & \thead{Local\\Choice} & \thead{Fixed\\Choice} & \thead{Local\\Saliency} \\
\hline
AUC
&$0.843$ 
& $0.835$ 
& $0.838$
& $0.804$ \\
\hline
NSS
&$1.48$ 
& $1.44$ 
& $1.41$
& $1.05$ \\
\noalign{\smallskip} \hline \noalign{\smallskip}
\end{tabular}
\caption{{\bf AUC and NSS of the different model variants.}
The full model performs better than the Local and Fixed Choice models. The Local Saliency Model performs significantly worse than the other model variants.}
\label{tab:roc_nss}
\end{table}

\section*{Discussion}
The current study proposed and analyzed a mathematical model of fixation selection, motivated by the Local and Global Attention modes that were suggested previously as a mechanism driving eye movements in natural scene-viewing tasks \cite{frost1976different,unema2005time,helmert2005two,tatler2008systematic}. We constructed a generative scan path model based on a small set of assumption. Using Bayesian \cite{mackay2003information} inference we fit the model to experimental data. By doing so, we continue the line of work of using generative likelihood based models for scan path generation \cite{Schuett2017}. Importantly, we use recent developments in Bayesian statistics to construct more efficient parameter inference algorithms. 

A different approach uses of deep neural networks for scan path modeling \cite{shao2017scanpath,kuemmerer2019deepgaze3}. One of the downsides of this approach is its reliance on large amounts of data, which precludes the study of interindividual differences. Thus, by using a hypothesis--based model, which requires only a relatively small number of parameters, we can fit individual models for each experimental subject and capture inter--subject variability. 

We demonstrate how our model captures the saccade amplitude both at the population and the individual level. Whereas two of the competing models perform equally well in terms of the coefficient of determination fitted to the mean and standard deviation of the individual subjects' saccade lengths, the advantage of the full model is demonstrated when looking at the autocorrelation of the saccade length. This analysis takes into account not only the individual saccades but also the dynamics of the entire scan path. These results emphasize the importance of the information about the saliency of the previously fixated location, when deciding on the next fixation location, as described in Eq.~\ref{eq:f}. Our model generates the typical behavior of a short saccade after a long one, or vice versa, which results in the observed anti-correlation of the length of subsequent saccades.

To further quantify the model performance we calculated the AUC and NSS scores of the different variants of the model. The full model and the Local and Fixed Choice variants performed similarly well, and the full model achieved slightly higher scores than the other two. This result indicates that these quantitative scores may not be enough to evaluate how well a scan path model fits the data. Rather than relying only on information-theoretic measures such as AUC and NSS, a more careful investigation of the full results suggests that the saccade behavior is specifically characterized by its autocorrelation function of the saccades amplitudes.

Next, we will discuss the limitations of the model. As described above, the Local and Global Attention Model successfully captures the experimental saccade amplitudes both at the population level and the subject level. Another spatial aspect of saccades is saccade direction. Our model captures only the tendency to perform horizontal saccades, but not the tendency to perform vertical saccades. This is expected from the construction of the model.

As the full model has information regarding the saliency of the previous fixation location, but not of the location itself, in its current form the model does not capture the change in saccade direction (i.e., the saccade direction relative to the previous saccade). The relative saccade direction is important for modeling known phenomena such as visual persistence or saccadic momentum \cite{ritter1976evidence,breitmeyer1982existence,wilming2013saccadic,luke2014dissociating}. As with the vertical preferred saccade direction, the model's inability to capture the relative saccade direction stems from the choice of Gaussian functions in the local and global attentional states. Our model could be extended to account for these tendencies by a mixture of Gaussians. Each Gaussian component would be designed to capture different directional tendencies, rather than capturing only one tendency as in the current version of the model. For example, one Gaussian can be aligned in the direction of the previous saccade, to account for visual persistence.

Other limitations of the model stem from the choice of a second order Markov process. Due to this choice the model is almost memory-less and cannot capture known phenomena in scene viewing which span multiple saccades. 
%One example of such a phenomenon is the well--known inhibition of return \cite{handy1999promoting,klein2000inhibition}. 
Incorporating longer history is not straightforward in our model. A heuristic approach could be including dynamics in the saliency map. 
%Currently we use a constant saliency map, but the model could be adapted to use an evolving saliency map accounting for effects such as inhibition of return.

Finally, our mathematical model does not account for fixation duration in scene viewing, which of course play an important role in eye-movement control \cite{henderson2003human,nuthmann2010crisp,Laubrock2013,Tatler2017}. So far most of the modeling attempts of scene viewing addressed either the spatial or the temporal aspects of scene viewing. Indeed, some models use temporal dynamics but they do not attempt to learn these dynamics from the data and use a heuristic--based approach. While fixation duration modeling is outside the scope of this work, we nonetheless consider the integration of temporal and spatial aspects an exciting new research direction.

\section*{Supporting information}

% Include only the SI item label in the paragraph heading. Use the \nameref{label} command to cite SI items in the text.

\paragraph*{S1 Appendix.}
\label{S1_Appendix}
{\bf Parameter inference.} In this appendix you can find the technical details of the Gibbs sampler described in the Methods section. It includes a derivation of the full likelihood, details regarding the augmentation schemes and the derivation of the conditional distributions.

\section*{Acknowledgments}
This research has been funded by Deutsche Forschungsgemeinschaft (DFG, German Research Foundation) - SFB 1294/1 - 318763901.

\nolinenumbers

% Either type in your references using
% \begin{thebibliography}{}
% Text
% \end{thebibliography}
%
% or
%
% Compile your BiBTeX database using our plos2015.bst
% style file and paste the contents of your .bbl file
% here. See http://journals.plos.org/plosone/s/latex for 

%\bibliography{bibliography2.bib}
%\printbibliography

%\bibliography{bibiliography.bib}

\newpage
\section*{Supplementary Material}
\subsection*{Likelihood augmentation}
In the Methods section we defined the likelihood function of the model and the prior distributions over the model parameters. Here, we present the details of the inference process for the estimation of the posterior distribution.

The full likelihood function of the model is
\begin{align}
\label{eq:qug_gamma}
    p(Z, \Gamma| \Theta) = p(z_1)p(z_2) \prod_{t=3}^{T} p_{\text{exploit}}\left(z_t|z_{t-1} \right)^{\gamma_{t}} \: p_{\text{explore}}\left(z_t|z_{t-1} \right)^ {1 - \gamma_{t}} p\left(\gamma_t \right) 
  \end{align}
 with
\begin{align*}
    p_{\text{exploit}}\left(z_t|z_{t-1} \right) &= \dfrac{n\left(z_t ; z_{t-1}, \epsilon \right)}{\sum_{z'} n\left(z' ; z_{t-1}, \epsilon \right)} \\
    p_{\text{explore}}\left(z_t|z_{t-1} \right) &= \dfrac{\max\left(s\left(z_t\right) n\left(z_t ; z_{t-1}, \xi \right) - n\left(z_t ; z_{t-1}, \epsilon \right), 0\right)}{\sum_{z'}\max\left(s\left(z'\right) n\left(z' ; z_{t-1}, \xi \right) - n\left(z' ; z_{t-1}, \epsilon \right), 0\right)}
\end{align*}
where
\begin{align*}
    \gamma_{t} \sim \text{Bern}\left(\rho_{t} \right) & = \: \text{Bern}\left(\sigma \left(f\left( s\right) \right) \right) \\ 
    & = \left(\dfrac{1}{1 + \exp \left( - f\left(s\right) \right)}\right)^{\gamma_t} \: \left(\dfrac{\exp \left(-f(s)\right)}{1 + \exp \left(-f\left(s\right) \right)}  \right)^{1 - \gamma_t} \\
    &=\dfrac{1}{1 + \exp \left(-f\left(s\right) \right)} \exp\left(-f\left(s\right) \right) ^ {1- \gamma_t} = \dfrac{2\exp \left(\left(\gamma_t  - \frac{1}{2} \right) \frac{f\left( s\right)}{2} \right)}{\cosh \left(\frac{f\left(s \right)}{2} \right)}
\end{align*}
and
\begin{align}
    f\left(s \right) &= b\left(\dfrac{s_{t-1}}{s_{t-2}} - s^o \right).
\end{align}

We choose the following prior distributions
\begin{align*}
    \epsilon_{x/y} &\sim \text{IG}\left(\alpha_{\epsilon_{x/y}}, \beta_{\epsilon_{x/y}} \right) \\
    \xi_{x/y} &\sim \text{IG}\left(\alpha_{\xi_{x/y}}, \beta_{\xi_{x/y}} \right) \\
    b &\sim \mathcal{N}\left(\mu_b, \sigma_b \right) \\
    s^o &\sim \mathcal{N}\left(\mu_{s^o}, \sigma_{s^o} \right).
\end{align*}

We wish to derive a Gibbs sampler to estimate the posterior distribution. To do so we need to derive the conditional distribution for each of the model parameters. This is not possible in the current form of the posterior distribution and we use an augmentation technique. The augmentation process described below results in an exponential quadratic form in $b$ and $s^o$ which allows for simple sampling from their respective conditional distributions.

In the process of data augmentation one introduces a set of auxiliary variables to make the model conditionally conjugated \cite{tanner1987calculation}. In this work we augment the model with Poly\'{a}-gamma variables $W$. We add one $w_t$ for each data point $z_t$. The Poly\'{a}-Gamma variables are defined in the following way \cite{polson2013bayesian,choi2013polya}.

The random variable $w \sim PG(1,0)$ is defined by its moment generating function:
\begin{align}
\label{eq:pg_moments}
\mathbb{E}\left(\exp(-tw)\right) =
\cosh^{-1}\left(\sqrt{\dfrac{t}{2}} \right).
\end{align}

We set $t=\frac{f\left( s\right)^2}{2}$ and apply Equation \ref{eq:pg_moments} to Equation \ref{eq:qug_gamma} which results in the following simplified posterior for the augmented model
\begin{align}
\label{eq:posterior_simp}
\nonumber
p\left(\Theta, W, \Gamma | Z\right) \propto \, & p\left( z_1\right)p\left( z_2\right)\prod_{t=3}^{T} p_{\text{exploit}}\left( z_t | z_{t-1} \right)^{\gamma_t} p_{\text{explore}}\left( z_t | z_{t-1} \right)^{1-\gamma_t} \times \\ 
&\exp \left(-\dfrac{f\left(s\right)^2}{2}w_t + \left(\gamma_t- \frac{1}{2}\right)f\left(s\right) \right)  p\left(w;1,0\right) p\left(\Theta \right)
\end{align}
with:
\begin{align}
    \Theta &= \{b, s^o, \epsilon, \xi \} \\
    p \left(\Theta \right) &= p(b)p(s^o)p(\epsilon)p\left(\xi\right).
\end{align}

We use the posterior distribution above in the Gibbs sampler and we next describe how to sample from the full conditionals of $W, \Gamma$ and the model parameters $\Theta$.

\subsection*{The Conditionals}

\subsubsection*{$w_t$}
It is not immediately clear how to sample from the conditional distribution for $w_t$. A helpful step is to identify the factors which depend on $w_t$ in Equation \ref{eq:posterior_simp} as part of the exponential tilted distribution of $w \sim PG(1,0)$. $w \sim PG(1,c)$ is the exponential tilting of $w \sim PG(1,0)$, and using the moment generating function introduced in Equation \ref{eq:pg_moments}:
\begin{align}\label{eq:pg_1_c}
p(w;1,c) = \dfrac{\exp(-\frac{c^2}{2}w)p(w|1,0)}{\mathbb{E}\left(\exp \left(-\frac{c^2}{2}w\right) \right)} = \cosh\left(\frac{c}{2}\right)\exp\big(-\frac{c^2w}{2}\big)p\left(w;1,0\right).
\end{align}
Setting $c = f\left(s\right)$ for each $w_t$ we conclude that to sample from the conditional distribution for $w_t$ we need to sample from $w \sim PG\left(1, f\left(s\right)\right)$. For the details of the sampling process see \cite{windle2014sampling}. In this work we used the Python implementation of the sampler described by Windel et al. called "pypolyagamma" and is available under 

\noindent https://github.com/slinderman/pypolyagamma .

\subsubsection*{$\gamma_t$}
To sample $\gamma_t$ we calculate the Bayes Factor and sample from $\text{Bern}(\gamma_t; \rho_t)$ with:
\begin{align}
\rho_t &= \dfrac{\sigma\left(f\left(s\right)\right)\: p\left(z_t | \gamma_t = 1\right)}{\sigma\left(f\left(s\right)\right)\: p\left(z_t | \gamma_t = 1\right) + \left(1 - \sigma\left(f\left(s\right)\right) \right)\: p\left(z_t | \gamma_t = 0\right)} \\ \nonumber &=\dfrac{\sigma\left(f\left(s\right)\right)}{\sigma\left(f\left(s\right)\right) + \mathrm{BF}\left(1- \sigma\left(f\left(s\right)\right)\right)}
\end{align}
and $\mathrm{BF}$ defined as:
\begin{align}
\mathrm{BF} = \dfrac{p\left(z_t|\gamma_t = 0\right)}{p\left(z_t|\gamma_t = 1\right)} = \dfrac{p_{\text{explore}}\left(z_t|z_{t-1} \right)}{p_{\text{exploit}}\left(z_t|z_{t-1} \right)}
\end{align}

\subsubsection*{$b$ and $s^o$}
Taking into account the definition of $f\left( s\right)$ we see that the parameters $b$ and $s^o$ appear in linear and quadratic forms in the arguments of the exponents in Equation (\ref{eq:posterior_simp}). Given the Gaussian prior distributions we chose, the conditional distributions of these parameters are also Gaussian
\begin{align}
p(b|Z, W, \Gamma, s^o, \epsilon, \xi) &= n\left( b;m_b, s_b\right)
\end{align}
with mean and variance
\begin{align}
m_b &= \sum_{t=1}^{T}\left(\gamma_t - \frac{1}{2}\right) \left(\frac{s_{t-1}}{s_{t-2}} - s^o \right) + \frac{\mu_b}{\sigma_b} s_b \\
s_b &= \dfrac{\sigma_b}{1 + \sigma_b \sum_{t=1}^{T}w_t \left(\frac{s_{t-1}}{s_{t-2}} - s^o\right)^2}.
\end{align}
Similarly
\begin{align}
     p(s^o|Z, W, \Gamma, b, \epsilon, \xi) &= n\left( s^o;m_{s^o}, s_{s^o}\right)
\end{align}
with mean and variance
\begin{align}
m_{s^o} &= \left( \sum_{t=1}^{T} \left(b^2 w_t \frac{s_{t-1}}{s_{t-2}} -b\left(\gamma_t - \frac{1}{2}\right)\right) + \frac{\mu_{s^o}}{\sigma_{s^o}}\right) s_{s^o} \\
s_{s^o} &= \dfrac{\sigma_{s^o}}{1 + \sigma_{s^o}b^2 \sum_{t=1}^{T}w_t }.
\end{align}

\subsubsection*{$\epsilon$ and $\xi$}
Due to the complex form of $p_{\text{explore}}$, we do not have a closed form for the conditional distributions of $\epsilon$ and $\xi$ from which we can sample. As an estimate we draw a sample using the Hamiltonian Monte Carlo (HMC) algorithm. The HMC algorithm requires an energy function and its derivative. In our case we use the negative log of the model likelihood as the energy function. Rather than calculating the derivative of the log likelihood analytically we use automatic differentiation \cite{griewank1989automatic}. Specifically we use the Python Autograd package. We further tuned the step size and number of steps parameters of the leapfrog algorithm to achieve an acceptance rate of $100\%$. Specifically we used a step size of $0.03$ for $\epsilon$ and $0.5$ for $\xi$ with eight leapfrog iterations for both.

\end{document}